\documentclass[
onecolumn;
superscriptaddress,
showpacs,
preprintnumbers,
amsmath,
amssymb,
aps,
prd,
]{revtex4-2}

\usepackage{graphicx}
\usepackage{dcolumn}
\usepackage{bm}

\begin{document}

\title{A nontopological soliton with a dipole chromomagnetic field}

\author{A.~Yu.~Loginov}

\email{a.yu.loginov@tusur.ru}

\affiliation{Laboratory of Applied Mathematics and Theoretical Physics, Tomsk State
           University of Control Systems and Radioelectronics, 634050 Tomsk, Russia}


\begin{abstract}
A non-Abelian  gauge  model  with  a  complex   isovector  scalar  field  and a
sixth-order self-interaction potential is considered.
It   is   shown   that   it    has    a    nontopological    soliton  solution.
The features of this  soliton  include  a  monopole-like  core  surrounded by a
Q-ball-like shell, the  existence of radially  excited states, and a long-range
dipole chromomagnetic field.
The properties  of  the  soliton  are   studied using  analytical and numerical
methods.
In particular, the  asymptotic  dependencies   of  the  energy  and the Noether
charge on  a   phase   frequency   are   obtained   for   two  extreme regimes.
It is also found that in  these two extreme regimes,  the chromomagnetic dipole
moment  of   the    soliton    is     proportional    to    its   linear  size.
\end{abstract}

\maketitle

\section{Introduction}                                            \label{sec:I}

Topological    and    nontopological    solitons    \cite{Manton,   E_Weinberg,
 lee_pang_1992} play an  important  role  in  mathematical physics, high-energy
physics, condensed matter physics, cosmology, and hydrodynamics.
The    't Hooft-Polyakov     monopole     \cite{hooft_74,  polyakov_74}   is  a
three-dimensional  topological  soliton   of  exceptional  importance  in field
theory.
A  classical  model  of  field  theory  that  admits   the   existence  of  the
't Hooft-Polyakov monopole is the $SU(2)$ Georgi-Glashow  model \cite{GG_1972}.
It includes  a real isovector scalar  field  $\boldsymbol{\phi}$ that interacts
minimally with a non-Abelian gauge field.
The fourth-order self-interaction potential of this model has a two-dimensional
sphere $\left\vert \boldsymbol{\phi} \right\vert  = \phi_{\text{vac}}$  as  the
true vacuum manifold.
The potential has no  false  vacuum  points, and, in particular, the zero point
$\boldsymbol{\phi} = 0$ is its local maximum.

The $SU(2)$ Georgi-Glashow   model    can    be    modified   in  several ways.
For example, the  self-interaction  potential  can  be  changed so that it will
have both a global minimum (true vacuum) and  a  local  minimum (false vacuum).
In particular, the  potential  of  the  model  considered  in \cite{kumar_2010,
 paranjape_2024} has the zero point $\boldsymbol{\phi} = 0$ as the true vacuum,
and  the   two-dimensional  sphere  $\left\vert \boldsymbol{\phi} \right\vert =
\phi_{\text{vac}}$ as the false vacuum manifold.
This model possesses metastable monopole solutions, which  interpolate  between
the true and false vacua.
Monopole solutions of  this  type  can significantly increase the decay rate of
the false     vacuum     \cite{kumar_2010, paranjape_2024, steinhardt_prd_1981,
 steinhardt_npb_1981, hosotani_prd_1983, agrawal_2022}.

In contrast to  \cite{kumar_2010, paranjape_2024},  the  potential of the model
considered in \cite{kim_plb_1999}  has  the  two-dimensional sphere $\left\vert
\boldsymbol{\phi}\right\vert = \phi_{\text{vac}}$  as  the true vacuum manifold
and the zero   point    $\boldsymbol{\phi}   =  0$    as    the  false  vacuum.
It has been shown in \cite{kim_plb_1999} that this  model has a solution, which
consists of a mono- pole-like core surrounded by  a region (bubble) of the true
vacuum.
Being a stationary point  of  an  Euclidean action, this gauged monopole-bubble
contributes to the decay amplitude of  the  false  vacuum  in the limit of high
temperature.

Another way of modification  is  to  add  a  new  field  to the original fields
of the $SU(2)$ Georgi-Glashow model.
In particular, the model considered in \cite{bai_2022}  contains an  additional
complex  scalar  field  interacting   with   the  real isovector   scalar field
$\boldsymbol{\phi}$ through a fourth-order potential term.
It has been shown  in  \cite{bai_2022}  that  this  model  has  a solution that
possesses both topological and  nontopological  charges, and therefore combines
properties of topological and nontopological solitons.

A classical  example  of  a  solution,  which   is   a  stationary point  of an
Euclidean action is the bounce.
A solution of this  type  was  first  described  in  \cite{coleman_1977} in the
framework of  the  model  of  a self-interacting real scalar field in Euclidean
space.
The bounce  determines  the   decay   amplitude   of   a  false  vacuum  in the
semiclassical approximation.
There is another solution of field theory called the Q-ball \cite{coleman_1985},
which is closely related to the bounce solution \cite{coleman_1977}.
The Q-ball  is  a  time-dependent  nontopological  soliton  of  the  model of a
self-interacting complex scalar field in Minkowski space.
The Q-ball is  a  stationary  point of the energy functional at a fixed Noether
charge.
The complex scalar field of  the  model results in the Noether charge, which is
necessary for the existence of the Q-ball.

In this paper, we consider a non-Abelian gauge model  with  a complex isovector
scalar field and a sixth-order self-interaction potential.
This   model    allows     the     existence    of  a  nontopological  soliton.
This soliton  is  related  to  the  gauged  monopole-bubble \cite{kim_plb_1999}
similar to how the Q-ball and the bounce are related.
The nontopological  soliton   of  the  model  consists  of a monopole-like core
surrounded by a shell of the Q-ball type.
Like the usual Q-ball, this shell can  be  either  in  the ground state or in a
radially excited state.
The energy and the Noether charge  of  the soliton depend on a phase frequency,
which determines the  time  dependence  of  the complex isovector scalar field.
A characteristic feature of the soliton is its  long-range chromomagnetic field
of the dipole type.
At the  same  time,   the   soliton   does  not  possess  any  electric  field.

The model under consideration is  a  modification of the $SU(2)$ Georgi-Glashow
model.
The  modification consists in complexification  of  the  Higgs isovector scalar
field and replacement of  the  original  renormalizable  fourth-order potential
with an effective sixth-order potential.
We note in this regard  that  potentials  of  field  models change shape during
phase transitions that occured in the early Universe,  and that a complex Higgs
isotriplet   exists    in    some   extensions   \cite{gunion_1990, rizzo_1991,
 espinosa_1992, dey_2009} of the Standard Model.

Q-balls   play   an   important  role   in   various   cosmological  scenarios.
Specifically, Q-balls can be produced efficiently in the Affleck–Dine mechanism
\cite{affleck_1985}    and   could   be   responsible   for   baryon  asymmetry
\cite{enqvist_1998, enqvist_1999, kasuya_2000}       and       dark      matter
\cite{kusenko_1998, kusenko_2009}.
This also applies to  Q-ball-like  solitons, since  their  main  component is a
scalar field condensate.
The model considered  in  this  paper  was  chosen  because  it is the simplest
non-Abelian gauge model in which a Q-ball-like soliton exists.
In addition, this soliton has a  long-range  chromomagnetic  field, which is an
additional motivation for research.

This  paper  is  structured as follows.
In Sec.~\ref{sec:II}, we describe briefly the Lagrangian, symmetries, and field
equations of the model under consideration.
In Sec.~\ref{sec:III}, the general  properties  of  the  nontopological soliton
are considered and discussed.
In particular, we derive two basic relations  which determine the properties of
the soliton.
In Sec.~\ref{sec:IV}, we study the  behavior  of  the nontopological soliton in
two (thin-wall and thick-wall) extreme regimes.
In Sec.~\ref{sec:V},    we    present    and    discuss    numerical   results.
In the final section, we  briefly summarize the results obtained in this paper.

Throughout the paper, the natural units $c = 1$ and $\hbar = 1$ are used.

\section{The Lagrangian and field equations of the model}        \label{sec:II}

The Lagrangian density of the model considered here has the form
\begin{equation}
\mathcal{L}=-\frac{1}{4}F_{\mu \nu }^{a}F^{a\,\mu \nu }+D_{\mu }\phi
^{a}D^{\mu}\phi^{a\ast}-V\left(\phi^{a}\phi^{a\ast}\right),        \label{II:1}                                                 \end{equation}
where
\begin{equation}
F_{\mu \nu }^{a}=\partial _{\mu }A_{\nu }^{a}-\partial _{\nu }A_{\mu
}^{a}-g\epsilon^{abc}A_{\mu }^{b}A_{\nu }^{c}                     \label{II:2}
\end{equation}
is the non-Abelian field strength,
\begin{equation}
D_{\mu}\phi^{a}=\partial_{\mu}\phi^{a}
-g\epsilon^{abc}A_{\mu}^{b}\phi^{c}                                \label{II:3}
\end{equation}
is  the  covariant   derivative   of   the   complex   isovector   scalar field
$\boldsymbol{\phi}$, and
\begin{equation}
V\left( \phi ^{a}\phi ^{a\ast }\right)  = \frac{m^{2}\phi^{a}\phi^{a\ast}
}{1+\varepsilon^{2}}\left[\left(1-\frac{\phi^{a}\phi^{a\ast}}{v^{2}}\right)
^{2}+\varepsilon ^{2}\right]                                       \label{II:4}
\end{equation}
is the sixth-order self-interaction  potential.
This form of writing  the  potential  was  used  in \cite{lee_pang_1992} in the
context of Q-balls; its advantage  is  that  it  has a zero  global  minimum at
$\boldsymbol{\phi}  =  0$  for  any  values  of  the  parameters  $m$, $v$, and
$\varepsilon$.
Hence, the  $SU(2)$   gauge   symmetry   of   the  model (\ref{II:1}) cannot be
spontaneously broken.

Using standard methods  of  field  theory, we obtain the field equations of the
model: 
\begin{eqnarray}
D_{\nu }F^{a\,\mu \nu }+g\epsilon^{abc}\left[ \phi ^{b}\left( D^{\mu }\phi
^{c}\right) ^{\ast }+\phi ^{b\ast }D^{\mu }\phi ^{c}\right] &=&0, \label{II:5a}
 \\
D_{\mu }D^{\mu }\phi ^{a}+V^{\prime }\left( \phi ^{a}\phi ^{a\ast }\right)
\phi ^{a} &=&0,                                                   \label{II:5b}
\end{eqnarray}
where
\begin{equation}
V^{\prime }\left( \phi ^{a}\phi ^{a\ast }\right) =m^{2}-\frac{4m^{2}\phi
^{a}\phi ^{a\ast }}{v^{2}\left( 1+\varepsilon ^{2}\right) }+\frac{
3m^{2}\left( \phi ^{a}\phi ^{a\ast }\right) ^{2}}{v^{4}\left( 1+\varepsilon
^{2}\right) }
\end{equation}
is the   derivative    of    the    potential   (\ref{II:4})   with  respect to
the sesquilinear combination $\phi^{a}\phi^{a\ast}$.
We  will  also  need  the  expression  of  the  energy-momentum  tensor  of the
model,
\begin{equation}
T_{\mu \nu }=-F_{\mu \rho }^{a}F_{\nu }^{a\,\rho }+D_{\mu }\phi ^{a}D_{\nu
}\phi ^{a\ast }+D_{\nu }\phi ^{a}D_{\mu }\phi ^{a\ast }+\eta _{\mu \nu }
\left[ \frac{1}{4}F_{\rho \tau }^{a}F^{a\,\rho \tau }-D_{\rho }\phi
^{a}D^{\rho }\phi ^{a\ast }+V\left( \phi ^{a}\phi ^{a\ast }\right) \right].
                                                                   \label{II:6}
\end{equation}

Besides the local gauge $SU(2)$ transformations, the Lagrangian (\ref{II:1}) is
also invariant under the global $U(1)$ transformations
\begin{equation}
\phi ^{a}\rightarrow e^{i\chi }\phi ^{a},\;\phi ^{a\ast}\rightarrow
e^{-i\chi}\phi^{a\ast}.                                            \label{II:7}
\end{equation}
The corresponding conserved Noether current is
\begin{equation}
j^{\mu }=i\left( \phi ^{a}D^{\mu }\phi ^{a\ast }-\phi ^{a\ast }D^{\mu }\phi
^{a}\right).                                                       \label{II:8}
\end{equation}
Note that the current (\ref{II:8})  is  invariant  under  both  the local gauge
$SU(2)$ transformations and the global $U(1)$ transformations.

The model (\ref{II:1}) differs from the well-known $SU(2)$ Georgi-Glashow model
in several important respects.
The  model  (\ref{II:1})  contains   the   complex   isovector   scalar   field
$\boldsymbol{\phi}$, whereas  the  $SU(2)$ Georgi-Glashow model contains a real
isovector scalar field.
Further, the  sixth-order  potential  (\ref{II:4})  has  a  zero global minimum
at the zero point $\boldsymbol{\phi} = 0$,  and  therefore the classical vacuum
does not break the gauge $SU(2)$ symmetry.
In contrast, the fourth-order potential of  the  Georgi-Glashow model reaches a
zero minimum on  a  two-dimensional sphere $\left\vert \boldsymbol{\phi} \right
\vert = \phi_{\text{vac}}$ resulting  in  a  spontaneous  breaking of the gauge
$SU(2)$ symmetry.

\section{The soliton solution and its properties}               \label{sec:III}

To find a soliton solution of  the  model (\ref{II:1}), we shall use a modified
't Hooft-Polyakov ansatz
\begin{eqnarray}
A^{a 0} &=&vj\left( r\right) n^{a},                              \label{III:8a}
 \\
A^{a i} &=&\epsilon ^{aim}n^{m}\frac{1-u\left( r\right) }{gr},   \label{III:8b}
 \\
\phi^{a} &=&2^{-1/2}vh\left( r\right) e^{i\omega t}n^{a},        \label{III:8c}
\end{eqnarray}
where $n^{a} = x^{a}/r$.
The modification consists  in  the  time  dependence $\propto e^{i\omega t}$ of
the complex isovector scalar field in Eq.~(\ref{III:8c}).
This   time   dependence   is  typical  for  nontopological  soliton  solutions
\cite{lee_pang_1992, fried_1976, coleman_1985}.
Substituting   Eqs.~(\ref{III:8a})--(\ref{III:8c})   into  the  field equations
(\ref{II:5a})--(\ref{II:5b}), we  obtain  a  system  of  nonlinear differential
equations for the ansatz functions:
\begin{eqnarray}
j^{\prime \prime }\left( r\right) +\frac{2}{r}j^{\prime }\left( r\right) -
\frac{2}{r^{2}}u(r)^{2}j(r) &=&0,                                \label{III:9a}
 \\
u^{\prime \prime }\left( r\right) -\frac{1}{r^{2}}u(r)\left( u(r)^{2}-1\right)
-\frac{m_{V}^{2}}{2}\left( h(r)^{2}-j(r)^{2}\right) u(r) &=&0,   \label{III:9b}
 \\
h^{\prime \prime }\left( r\right) +\frac{2}{r}h^{\prime }\left( r\right) -
\frac{2}{r^{2}}u(r)^{2}h(r)+m^{2}\partial \tilde{U}_{\omega }\left(
h(r)\right) /\partial h &=&0,                                    \label{III:9c}
\end{eqnarray}
where
\begin{equation}
\tilde{U}_{\omega}\left(h\right)=-\frac{1}{2}\left(\frac{m_{\omega}^{2}}
{m^{2}}h^{2} - \frac{h^{4}}{1+\varepsilon ^{2}}+\frac{1}{4}\frac{h^{6}}
{1+\varepsilon ^{2}}\right)                                      \label{III:10}
\end{equation}
is a dimensionless effective potential, and  $m_{V}=\sqrt{2}gv$ and $m_{\omega}
= (m^{2} - \omega^{2})^{1/2}$ are the mass parameters.

It follows  from   Eqs.~(\ref{II:4})  and  (\ref{III:8c})  that  the  potential
(\ref{II:4}) vanishes only at zeros of the ansatz function $h(r)$.
Then, from the condition  of the finiteness of the soliton's energy, it follows
that $\underset{r\rightarrow \infty}{\lim}h\left(r\right) = 0$.
Hence, we can  neglect  the terms $-m_{V}^{2}h^{2}u/2$ and $-u(u^{2}- 1)/r^{2}$
as $r \rightarrow \infty$,  and  Eq.~(\ref{III:9b}) takes  the  asymptotic form
\begin{equation}
u^{\prime \prime}\left(r\right) + \frac{m_{V}^{2}}{2}j(r)^{2}u(r) = 0.
                                                                 \label{III:11}
\end{equation}
Eq.~(\ref{III:11})  tells  us  that  if  $\underset{r \rightarrow \infty}{\lim}
j\left( r\right) \equiv j_{\infty} \neq 0$, then  the  ansatz  function  $u(r)$
will oscillate at large $r$, which is  incompatible  with the finiteness of the
energy of the soliton solution.
We thus conclude that the limiting value $j_{\infty} = 0$.

Further, Eq.~(\ref{III:8a}) tells us that the regularity of the solution at the
origin leads to another boundary condition $j(0) = 0$, so  that $j(r)$ vanishes
at these two boundary points.
To satisfy  these  homogeneous  boundary  conditions,  the  positive (negative)
ansatz function $j(r)$ must reach a global maximum (minimum) at some point $r =
\bar{r}$.
In other words,  the  relation  $\text{sgn}\left(j^{\prime\prime} \left(\bar{r}
\right)\right)=-\text{sgn}\left(j\left(\bar{r}\right)\right)$ must hold at this
point.
However, it follows from Eq.~(\ref{III:9a}) that $j^{\prime\prime}\left(\bar{r}
\right)=2u\left(\bar{r}\right)^{2}j\left(\bar{r}\right)/\bar{r}^{2}$, and hence
$\text{sgn}\left(j^{\prime\prime}\left(\bar{r}\right)\right) = \text{sgn}\left(
j\left(\bar{r}\right)\right)$.
It follows  from   the   contradiction   obtained   that  the  ansatz  function
\begin{equation}
j\left( r\right) = 0.                                            \label{III:12}
\end{equation}
Using Eqs.~(\ref{II:2}), (\ref{III:8a}), and (\ref{III:8b}), we obtain the
expression for the chromoelectric field $E^{ai}=F^{ai0}$
\begin{equation}
E^{a i}=-vj\left(r\right)u\left(r\right)\left(\delta^{ai}-n^{a}n^{i}\right)/r
-vj^{\prime }\left( r\right) n^{a}n^{i}.                        \label{III:12b}
\end{equation}
It follows from  Eqs.~(\ref{III:12}) and (\ref{III:12b}) that  a  finite energy
soliton solution cannot  possess a chromoelectric field.

Using   Eqs.~(\ref{II:6}),  (\ref{III:8b}),   and  (\ref{III:8c}),   we  obtain
the expression for the energy  of  the  soliton solution in terms of the ansatz
functions $u(r)$ and $h(r)$
\begin{equation}
E=4\pi v^{2}\int\limits_{0}^{\infty }\biggl[ \frac{\omega ^{2}}{2}h\left(
r\right) ^{2}+\frac{2u^{\prime }\left( r\right) ^{2}}{m_{V}^{2}r^{2}}+\frac{
\bigl( u\left( r\right) ^{2}-1\bigr) ^{2}}{m_{V}^{2}r^{4}}+\frac{1}{2}
h^{\prime }\left( r\right) ^{2}+\frac{h\left( r\right) ^{2}u\left( r\right)
^{2}}{r^{2}}+m^{2}\tilde{V}\left( h\left( r\right)\right) \biggr]r^{2}dr,
                                                                 \label{III:13}
\end{equation}
where the dimensionless self-interaction potential is
\begin{equation}
\tilde{V}\left( h\right)=\frac{1}{2}\left(h^{2}-\frac{h^{4}}{1+\varepsilon^{2}
}+\frac{1}{4}\frac{h^{6}}{1+\varepsilon^{2}}\right).             \label{III:14}
\end{equation}
Similarly,  we   obtain  the  expression  for  the  soliton's   Noether  charge
\begin{equation}
Q = 4\pi v^{2}\omega\int\limits_{0}^{\infty}h\left(r\right)^{2}r^{2}dr,
                                                                 \label{III:15}
\end{equation}
while the spatial components of the Noether current (\ref{II:8}) vanish for the
field configurations (\ref{III:8a})--(\ref{III:8c}).
We also need the expression for  the  soliton's  chromomagnetic field $B^{ai} =
-\epsilon ^{ijk}F^{ajk}/2$ in terms of the ansatz functions
\begin{equation}
B^{ai}=\frac{u^{\prime }\left( r\right) }{gr}\left( \delta
^{ai}-n^{a}n^{i}\right) -\frac{1-u\left( r\right) ^{2}}{gr^{2}}n^{a}n^{i}.
                                                                \label{III:15b}
\end{equation}

We now investigate the  asymptotics  of the ansatz functions $u(r)$ and $h(r)$.
Substituting the power  expansions  of  these functions  in Eqs.~(\ref{III:9b})
and  (\ref{III:9c}), we get the small $r$ asymptotics
\begin{eqnarray}
u\left( r\right)  & = &1+\frac{u_{2}}{2!}r^{2}+\frac{u_{4}}{4!}
r^{4}+O\left( r^{6}\right),                                     \label{III:16a}
 \\
h\left( r\right)  & = &h_{1}r+\frac{h_{3}}{3!}r^{3}+O\left( r^{5}\right),
                                                                \label{III:16b}
\end{eqnarray}
where the next-to-leading expansion coefficients are
\begin{eqnarray}
u_{4}&=&\frac{3}{5}\left(3u_{2}^{2}+2m_{V}^{2}h_{1}^{2}\right), \label{III:17a}
 \\
h_{3} &=&\frac{3}{5}h_{1}\left( m_{\omega }^{2}+2u_{2}\right).  \label{III:17b}
\end{eqnarray}
It follows from  Eqs.~(\ref{III:16a})--(\ref{III:16b})  that  $u(r)$ and $h(r)$
are even and odd functions of $r$, respectively.

Next, we    investigate     the    large    $r$     asymptotics    of   $u(r)$.
We first note that $u(r)$ tends to some constant $u_{\infty}$ as $r \rightarrow
\infty$, otherwise the energy (\ref{III:13}) would be infinite.
It follows that at large  $r$, the  ansatz  function  $u(r)$  can be written as
$u(r)=u_{\infty}+\delta \left( r\right)$, where $\delta \left( r\right) \ll 1$.
Furthermore,   Eq.~(\ref{III:9c})   tells   us   that   $h(r)$  tends  to  zero
exponentially as $r \rightarrow \infty$,  provided that $u\left(r\right)$ tends
to a constant $u_{\infty}$.
Taking  this  into account, we obtain the asymptotic form of Eq.~(\ref{III:9b})
in terms of $\delta(r)$
\begin{equation}
\delta ^{\prime \prime }\left( r\right) +\frac{1-3u_{\infty }^{2}}{r^{2}}
\delta \left( r\right) +\frac{u_{\infty }\left( 1-u_{\infty }^{2}\right)}
{r^{2}}=0,                                                       \label{III:18}
\end{equation}
where we neglect terms  nonlinear  in  $\delta$  because  of their smallness in
comparison with the linear term.

The solution to Eq.~(\ref{III:18}) is
\begin{equation}
\delta \left( r\right) =\frac{u_{\infty }\left( u_{\infty }^{2}-1\right)}
{1-3u_{\infty}^{2}}+b_{1}r^{\alpha_{1}}+b_{2} r^{\alpha _{2}}.   \label{III:19}
\end{equation}
where the exponents
\begin{equation}
\alpha _{1,2}=\frac{1}{2}\bigl( 1\pm i\sqrt{3\left( 1-4u_{\infty
}^{2}\right) }\bigr).                                            \label{III:20}
\end{equation}
In Eq.~(\ref{III:19}), the  inhomogeneous  term  must  vanish, since $\delta(r)
\rightarrow  0$  as  $r  \rightarrow  \infty$,  which  leads  to  the following
alternatives:
\begin{equation}
u_{\infty}=0,\;\alpha_{1,2}=\frac{1}{2}\bigl(1\pm i\sqrt{3}\bigr)\label{III:21}
\end{equation}
or
\begin{equation}
u_{\infty }=\pm 1,\;\alpha _{1}=-1,\;\alpha _{2}=2.              \label{III:22}
\end{equation}
The first alternative corresponds to an oscillating solution with an increasing
amplitude and is, therefore, unacceptable.
We are left with the second option in which we choose the upper sign, since the
regularity at the origin results in the boundary condition $u(0) = 1$.
Thus, we have the following large $r$ asymptotics of $u(r)$
\begin{equation}
u(r) \sim 1 + \frac{b}{r} + O\left(\frac{1}{r^{2}}\right),       \label{III:23}
\end{equation}
where $b$ is a constant.

Using Eq.~(\ref{III:9b}), it  is  easy to show that $u(r)$ cannot exceed unity,
since otherwise it would increase without bound.
It follows that in Eq.~(\ref{III:23}), $b$ is negative as well as   $u_{2}$  in
Eq.~(\ref{III:16a}).
Taking  into  account  that  the   limit  value  $u_{\infty}  =  1$  and  using
Eq.~(\ref{III:9c}), we  finally obtain  the  large  $r$  asymptotics of $h(r)$:
\begin{equation}
h\left( r\right) \propto  \frac{e^{-m_{\omega }r}}{m_{\omega }r}\left[1+\frac{1
}{m_{\omega }r} 
+O\left( \frac{1}{r^{2}}\right) \right].          \label{III:24}
\end{equation}
Comparing Eqs.~(\ref{III:23}) and (\ref{III:24}), we see that as $r \rightarrow
\infty$, $u(r)$ has long-range asymptotics $\propto r^{-1}$, whereas $h(r)$ has
short-range exponential asymptotics.

Combining Eqs.~(\ref{III:8b})  and  (\ref{III:23}),  we  obtain  the  large $r$
asymptotics of the gauge field
\begin{equation}
A^{a i}\sim -\epsilon ^{aim}n^{m}\frac{b}{gr^{2}}.               \label{III:25}
\end{equation}
Eq.~(\ref{III:25}) tells  us  that  the soliton has a long-range gauge field of
dipole type.
Indeed, Eq.~(\ref{III:25}) can be rewritten as
\begin{equation}
\mathbf{A}^{a}\sim \frac{1}{4\pi }
\frac{\mathbf{m}^{a}\!\times \mathbf{r}}{r^{3}},                 \label{III:26}
\end{equation}
where
\begin{equation}
m^{ai}=\frac{4\pi }{g}b\delta ^{ai}.                             \label{III:27}
\end{equation}
Eq.~(\ref{III:26}) coincides with the expression  for the vector potential of a
chromomagnetic dipole with the dipole moment (\ref{III:27}).
Eqs.~(\ref{III:15b}) and (\ref{III:23}) lead to the asymptotics of the soliton's
chromomagnetic field 
\begin{equation}
B^{ai}\sim -\frac{b}{gr^{3}}\left(\delta^{ai}-n^{a}n^{i}\right)+
\frac{2b}{gr^{3}} n^{a}n^{i},                                    \label{III:28}
\end{equation}
or, in vector form,
\begin{equation}
\mathbf{B}^{a} \sim \frac{1}{4\pi r^{3}}\left[3\mathbf{n}\left( \mathbf{m}
^{a}\!\!\cdot\!\mathbf{n}\right) -\mathbf{m}^{a}\right].         \label{III:29}
\end{equation}

The soliton's energy (\ref{III:13})  can  be  written  as the sum of four terms
\begin{equation}
E = E^{\left(T\right)}+E^{\left(B\right)}+E^{\left(G\right)}+
E^{\left(P\right)},                                              \label{III:30}
\end{equation}
where
\begin{equation}
E^{\left( T\right) }=4\pi v^{2}\int\limits_{0}^{\infty }\frac{\omega ^{2}}{2
}h\left( r\right) ^{2}r^{2}dr                                    \label{III:31}
\end{equation}
is the kinetic part of the energy,
\begin{equation}
E^{\left( B\right) }=\frac{4\pi }{g^{2}}\int\limits_{0}^{\infty }\left[
\frac{u^{\prime }\left( r\right) ^{2}}{r^{2}}+\frac{\bigl( u\left( r\right)
^{2}-1\bigr) ^{2}}{2r^{4}}\right] r^{2}dr                        \label{III:32}
\end{equation}
is the energy of the chromomagnetic field,
\begin{equation}
E^{\left( G\right) }=4\pi v^{2}\int\limits_{0}^{\infty }\left[ \frac{1}{2}
h^{\prime }\left( r\right) ^{2}+\frac{h\left( r\right) ^{2}u\left( r\right)
^{2}}{r^{2}}\right] r^{2}dr                                      \label{III:33}
\end{equation}
is the gradient part of the energy, and
\begin{equation}
E^{\left( P\right) }=4\pi v^{2}\int\limits_{0}^{\infty }
m^{2}\tilde{V}\left( h\left(r\right) \right) r^{2}dr             \label{III:34}
\end{equation}
is the potential part of the energy.
The Lagrangian $L = \int \mathcal{L}d^{3}x$  can  also be expressed as a linear
combination of these four terms:
\begin{equation}
L=E^{\left(T\right)}-E^{\left(B\right)}-E^{\left(G\right)}-
E^{\left(P\right)}.                                              \label{III:35}
\end{equation}
Furthermore, these four  terms are related by one more linear (virial) relation
\begin{equation}
3E^{\left(T\right)}+E^{\left(B\right)}-E^{\left(G\right)}-
3E^{\left(P\right)} = 0.                                         \label{III:36}
\end{equation}

To derive Eq.~(\ref{III:36}),  we  need  to  remember  that  the nontopological
soliton solution is  an  extremum of the action $S = \int\nolimits_{0}^{T} \int
\mathcal{L}d^{3}xdt$.
It follows from Eqs.~(\ref{III:31})--(\ref{III:35}) that the Lagrangian density
(\ref{II:1}) is time-independent, which implies that $S = L T$.
Hence, the soliton solution  is  also  an  extremum of the Lagrangian $L = \int
\mathcal{L}d^{3}x$.

After the scale transformation $r\rightarrow\varkappa r$ of the argument of the
ansatz functions $u(r)$ and $h(r)$ of the soliton  solution, the Lagrangian $L$
becomes a function of the scale parameter $\varkappa$.
Since the function $L (\varkappa)$ has a stationary point at $\varkappa=1$, its
derivative vanishes  at this point: $\left. dL/d\varkappa\right\vert_{\varkappa
= 1} = 0$.
It can easily be shown that  under  the  rescaling $r \rightarrow \varkappa r$,
$E^{\left( T\right) }\rightarrow \varkappa^{-3}E^{\left(T\right)}$,  $E^{\left(
B\right)}  \rightarrow  \varkappa   E^{\left( B\right)}$,  $E^{\left( G\right)}
\rightarrow    \varkappa^{-1}E^{\left( G\right) }$,  and   $E^{\left( P\right)}
\rightarrow \varkappa ^{-3}E^{\left( P\right)}$.
Using  these transformation  rules  and  Eqs.~(\ref{III:31}) -- (\ref{III:35}),
we  obtain the virial relation (\ref{III:36}).

Using Eqs.~(\ref{III:13}),  (\ref{III:15}),  and  (\ref{III:35}), it is easy to
show that the energy, the Lagrangian, and the Noether charge of the soliton are
connected by the linear relation
\begin{equation}
L = \omega Q - E.                                                \label{III:37}
\end{equation}
Since the   soliton   solution   is   an   extremum  of the Lagrangian $L$, the
following variational relation must hold
\begin{equation}
\delta L = \omega \delta Q - \delta E = 0.                       \label{III:38}
\end{equation}
Eq.~(\ref{III:38}) tells  us  that  the  soliton solution is an extremum of the
energy functional $E$ among the field configurations of the Noether charge $Q$.
Thus, the  soliton  solution  is  not  only  an  unconditional  extremum of the
Lagrangian, but also a conditional extremum of the energy.

In Eq.~(\ref{III:38}),  the  phase  frequency  $\omega$  plays  the role of the
Lagrange multiplier,  and  therefore  variations  of  the fields are arbitrary.
In particular, these  variations  can  link  two  infinitesimally close soliton
solutions.
In this case, Eq.~(\ref{III:38}) leads  to  the important differential relation
\begin{equation}
dE/dQ = \omega,                                                  \label{III:39}
\end{equation}
which relates  the  derivative  of  the  soliton's  energy  with respect to the
Noether charge with the phase frequency $\omega$.
Note that a similar differential relation also  holds  for electrically charged
topological solitons \cite{loginov_plb_2021, loginov_plb_2024}.

\section{Extreme regimes of the soliton} \label{sec:IV}

The  phase   frequency   $\omega$   of   the   complex  isovector  scalar field
$\boldsymbol{\phi}$  is  the most important  parameter; it determines the basic
properties of the soliton.
The phase    frequency    may     vary     only     in    a     limited region.
It follows from  Eq.~(\ref{III:24})  that  $\left\vert \omega \right\vert < m$.
Otherwise, the  parameter  $m_{\omega} = (m^{2} - \omega^{2})^{1/2}$  is purely
imaginary, the asymptotics (\ref{III:24}) is  oscillatory,  and  the  soliton's
energy (\ref{III:13}) is infinite.

Also, the magnitude of  the  phase  frequency must be greater than some minimum
value  $\left\vert  \omega \right\vert  > \omega_{\min}$,  because  the soliton
solution does not exist otherwise.
A qualitative analysis of  the  differential equation (\ref{III:9c}) based on a
mechanical  analogy  \cite{coleman_1977} shows that  in  order  for  a  soliton
solution  to  exist,  the   effective   potential  (\ref{III:10})  must  have a
non-negative maximum at some nonzero $h$.
It is easy to show that this is only possible if $\left\vert\omega\right\vert >
\omega _{\min }= m \varepsilon (1+\varepsilon ^{2})^{-1/2}$.
We conclude that the soliton may exist, provided that
\begin{equation}
m\varepsilon \left(1+\varepsilon^{2}\right)^{-1/2} < \left\vert \omega
\right\vert < m.                                                   \label{IV:1}
\end{equation}

\subsection{The thin-wall regime} \label{subsec:IVA}

The thin-wall regime corresponds  to  a  situation  in which $\left\vert \omega
\right\vert \rightarrow \omega_{\min}=m\varepsilon (1+\varepsilon^{2})^{-1/2}$.
An analysis based on the  mechanical  analogy  \cite{coleman_1977} reveals that
in this regime, the soliton can be divided into three regions.
In the inner region $r<R_{\text{c}}$,  we  approximate the ansatz functions by
the first terms of  their power expansions (\ref{III:16a}) and (\ref{III:16b})
\begin{eqnarray}
u\left( r\right)  &=&1-r^{2}R_{\text{c}}^{-2},                    \label{IV:2a}
 \\
h\left( r\right) &=&\bigl( 2^{1/2}-s \bigr) r R_{\text{c}}^{-1}.  \label{IV:2b}
\end{eqnarray}
In Eq.~(\ref{IV:2b}), $2^{1/2}$  is  the  value  of $h$  at which the effective
potential $\tilde{U}_{\omega_{\min}}(h)$  in  Eq.~(\ref{III:10}) reaches a zero
maximum, and the value of the parameter $s$ is assumed to be small.
Substituting Eqs.~(\ref{IV:2a})  and  (\ref{IV:2b}) into Eq.~(\ref{III:35}), we
obtain the contribution of the inner region to the Lagrangian
\begin{equation}
L_{1}=\frac{a_{-1}}{R_{\text{c}}}+a_{1}R_{\text{c}}+a_{3}R_{\text{c}}^{3},
                                                                   \label{IV:3}
\end{equation}
where the coefficients
\begin{eqnarray}
a_{-1} &=&-\frac{234}{35}\frac{\pi }{g^{2}},                      \label{IV:4a}
 \\
a_{1}&\approx &-\frac{34}{35}\pi\left( 2-2\sqrt{2}s\right) v^{2}, \label{IV:4b}
 \\
a_{3} &\approx &\frac{4}{315}\pi v^{2}\left( \frac{55 m^{2}}{1+\varepsilon
^{2}}-63m_{\omega }^{2}\right) + \frac{4}{105}\sqrt{2}\pi v^{2}\left( 21
m_{\omega }^{2} - \frac{25 m^{2}}{\left( 1+\varepsilon ^{2}\right)
}\right) s,                                                       \label{IV:4c}
\end{eqnarray}
and we hold the terms up to the first order in $s$.

In the  intermediate  region $R_{\text{c}} < r < R$, we  approximate the ansatz
functions by
\begin{eqnarray}
u(r) &=&0,                                                        \label{IV:5a}
 \\
h\left(r\right) &=&2^{1/2}-s.                                     \label{IV:5b}
\end{eqnarray}
Using Eqs.~(\ref{IV:5a}) and (\ref{IV:5b}),  we  obtain the contribution of the
intermediate region to the Lagrangian
\begin{equation}
L_{2}=\frac{4}{3}\pi v^{2}\left(R^{3}-R_{\text{c}}^{3}\right)\left(\Omega
^{2}-\sqrt{2}\Omega^{2}s-2^{-1}\mu ^{2}s^{2}\right),               \label{IV:6}
\end{equation}
where
\begin{equation}
\Omega^{2}=\omega^{2}-\omega_{\min}^{2}                            \label{IV:7}
\end{equation}
and
\begin{equation}
\mu ^{2}\approx -\left. d^{2}\tilde{U}_{\omega_{\text{min}}}\left( h\right)
/dh^{2}\right\vert_{h=\sqrt{2}}=4m^{2}\left(1+\varepsilon^{2}
\right)^{-1}.                                                      \label{IV:8}
\end{equation}

We    now     turn     to     the     third    (outer)    region     $r  >  R$.
We assume that $R$ is large enough to fulfill the conditions $m_{V}R \gg 1$ and
$m_{\omega}R \gg 1$.
Then we  can  neglect  the  terms  $\propto  r^{-1}$  and  $\propto  r^{-2}$ in
Eq.~(\ref{III:9c}), which allows us to lower its  order by one resulting in the
first-order equation
\begin{equation}
h^{\prime}=-m\sqrt{2\bigl\vert \tilde{U}_{\omega_{\min }}\left(h\right)
\bigr\vert}.                                                       \label{IV:9}
\end{equation}
Using Eqs.~(\ref{III:35}) and (\ref{IV:9}), it can be shown that in the leading
order in $R$, the  contribution  of  the  outer  region  to  the  Lagrangian is
\begin{equation}
L_{3}=-4\pi R^{2}T,                                               \label{IV:10}
\end{equation}
where the surface tension
\begin{equation}
T\approx mv^{2}\int\nolimits_{0}^{\sqrt{2}}
\sqrt{2\bigl\vert \tilde{U}_{\omega_{\min }}
\left( h\right) \bigr\vert }dh = \frac{mv^{2}}
{2\sqrt{1+\varepsilon ^{2}}}.                                     \label{IV:11}
\end{equation}
Note that the  contribution  of  the  magnetic  field  of  the  outer region to
$L_{3}$ is $\propto R^{-1}$, and  therefore  can  be  neglected in the limit of
large $R$.

The Lagrangian of the soliton  configuration is $L=L_{1}+L_{2}+L_{3}$, and is a
function  of  three   parameters:   $L   =   L\left(R_{\text{c}}, R, s\right)$.
The soliton solution  is  an  extremum  of  the Lagrangian, which leads us to a
system of three nonlinear equations
\begin{equation}
\partial_{R_{\text{c}}}L=0,\;\partial_{R}L=0,\;\partial _{s}L=0.  \label{IV:12}
\end{equation}
In the leading order in the small parameter $\Omega^{2} = \omega^{2} -\omega^{2
}_{\min}$, the solution of the system (\ref{IV:12}) is
\begin{equation}
R_{\text{c}}\approx \frac{\sqrt{51}}{4m}\left( 1+\varepsilon ^{2}\right)
^{1/2}\left[ \left( 1+\frac{1248}{289}\frac{m^{2}m_{V}^{-2}}{1+\varepsilon
^{2}}\right) ^{1/2}-1\right] ^{1/2},                              \label{IV:14}
\end{equation}
\begin{equation}
s\approx -\sqrt{2}\Omega ^{2}\mu ^{-2},                           \label{IV:15}
\end{equation}
and
\begin{equation}
R\approx \frac{2T}{\Omega^{2}\left(\Omega ^{2}+\mu^{2}\right)}
\frac{\mu^{2}}{v^{2}} \approx \frac{2T}{v^{2}\Omega^{2}} =
\frac{2T}{v^{2}\left(\omega ^{2}-\omega_{\min}^{2}\right)}.       \label{IV:16}
\end{equation}
Substituting   Eqs.~(\ref{IV:14}) -- (\ref{IV:16})    into    the    Lagrangian
$L\left(R_{\text{c}}, R,s\right)$ and keeping the higher order terms in $R$, we
obtain the Lagrangian  of  the  soliton  solution  as  a  function of the phase
frequency $\omega$
\begin{equation}
L\left(\omega\right)\approx-\frac{16\pi T^{3}\mu^{4}}{3v^{4}\Omega^{4}
\left(\Omega^{2} + \mu^{2}\right)^{2}},                           \label{IV:17}
\end{equation}
where $\Omega^{2} = \omega^{2} - \omega^{2}_{\min}$.

From Eqs.~(\ref{III:37}),  (\ref{III:39}),  and  (\ref{IV:17}), it follows that
the energy $E(Q)$  and  the  Lagrangian $L(\omega)$ of the soliton solution are
related  through  the  Legendre  transformation $E\left(Q\right)   =   \omega Q
- L\left( \omega \right)$.
Using  the  known  properties   of   the   Legendre  transformation,  we obtain
successively the  expressions  of  the  Noether  charge  and  the energy of the
soliton in the thin-wall approximation
\begin{equation}
Q\left( \omega \right) = \frac{dL}{d\omega }=\frac{64\pi T^{3}}{3v^{4}}\frac{
\omega }{\left( \omega ^{2}-\omega _{\min }^{2}\right)^{3}}       \label{IV:18}
\end{equation}
and $E\left( Q\right) =E\left( \omega \left( Q\right) \right)$, where
\begin{equation}
E\left( \omega \right) =\omega \frac{dL}{d\omega }-L\left( \omega \right) =
\frac{16 \pi T^{3}}{3v^{4}}\frac{5\omega ^{2} - \omega_{\min}^{2}}{\left(
\omega^{2}-\omega_{\min }^{2}\right)^{3}}                         \label{IV:19}
\end{equation}
and  the  dependence   $\omega(Q)$   is   determined   from  Eq.~(\ref{IV:18}).
Using Eqs.~(\ref{IV:18})  and  (\ref{IV:19}), we  find  that  in  the thin-wall
regime, the ratio $E/Q$ is
\begin{equation}
\frac{E}{Q}=\omega \left( \frac{5}{4}-\frac{\omega _{\min }^{2}}{4\omega ^{2}
}\right),                                                         \label{IV:20}
\end{equation}
and the derivative $dE/dQ$ is
\begin{equation}
\frac{dE}{dQ} = \frac{dE/d\omega}{dQ/d\omega} = \omega.           \label{IV:21}
\end{equation}
We see that both $E/\left\vert Q \right\vert$ and $dE/d\left\vert Q\right\vert$
tend to the  same  limit  $\omega_{\min} = m \varepsilon (1+\varepsilon ^{2})^{
-1/2}$ as $\left\vert \omega \right\vert \rightarrow \omega_{\min}$.

It follows from Eq.~(\ref{IV:14})  that  in  the  thin-wall  regime, the radius
$R_{\text{c}}$  of  the  inner  region  (monopole-like core) is finite and does
not depend on the phase frequency $\omega$.
In contrast, Eq.~(\ref{IV:16}) tells us that the radius $R$ of the intermediate
region of the Q-ball type  increases without bound as $\left\vert \omega \right
\vert \rightarrow \omega_{\min}$.
The energy and the  Noether  charge  of  the  soliton  are $\propto R^{3}$, and
therefore they also  increase without bound  as $\left\vert \omega \right \vert
\rightarrow \omega_{\min}$.
The effective thickness  of  the  outer  region and the surface tension $T$ are
practically independent  of  the phase frequency $\omega$ and can be considered
as constants.
Finally,  the  long-range  chromomagnetic  field (\ref{III:28})  of dipole type
extends far beyond the outer region of the soliton.

\subsection{The thick-wall regime} \label{subsec:IVB}

In the thick-wall regime, the magnitude of  the  phase frequency $\omega$ tends
to a maximum value equal to $m$.
In this regime, the effective potential (\ref{III:10})  has  a first-order zero
located at $h_{0}\approx m_{\omega}m^{-1}(1+\varepsilon^{2})^{1/2}$, where $m_{
\omega} = (m^{2} - \omega^{2})^{1/2}$.
We see that this zero is located  in a  small neighborhood of $h=0$, since $m_{
\omega}$ is small in the thick-wall regime.
The  analysis  of  the  system  (\ref{III:9b}) -- (\ref{III:9c})  based  on the
mechanical analogy \cite{coleman_1977}  shows  that  the  maximum  value of the
ansatz function $h(r)$ is of the order of $h_{0}$, and hence is small.
At the same time, the ansatz  function  $u(r)$ is not small and is of the order
of unity.
Finally,  Eq.~(\ref{III:24})  tells   us  that  the  dimensionless  combination
$m_{\omega} r$ is natural in the thick-wall regime.

In the light of the above,  we write the ansatz functions in the form $u\left(r
\right) =\tilde{u}\left(\varrho\right)$  and $h\left(r\right)=m_{\omega} m^{-1}
\tilde{h}\left( \varrho \right)$, where $\varrho = m_{\omega}r$.
Using these expressions  for  the ansatz functions, we find that the Lagrangian
can be written as
\begin{equation}
L=m_{\omega }\tilde{L}_{1}+m_{\omega }^{3}\tilde{L}_{3},          \label{IV:22}
\end{equation}
where
\begin{equation}
\tilde{L}_{1} = 4\pi v^{2}\int\limits_{0}^{\infty }\left[ -\frac{2\tilde{u}
^{\prime }\left( \varrho \right) ^{2}}{m_{V}^{2}\varrho ^{2}}-\frac{\left(
\tilde{u}\bigl( \varrho \right) ^{2}-1\bigr) ^{2}}{m_{V}^{2}\varrho ^{4}}
- \frac{1}{2m^{2}}\tilde{h}^{\prime }\left( \varrho \right) ^{2}-
\frac{\tilde{h}\left( \varrho \right) ^{2}\tilde{u}\left( \varrho \right)
^{2}}{m^{2}\varrho ^{2}}+\frac{1}{m^{2}}\tilde{W}
\bigl( \tilde{h}\left( \varrho \right) \bigr) \right] \varrho^{2}d\varrho,
                                                                  \label{IV:23}
\end{equation}
\begin{equation}
\tilde{W}\bigl( \tilde{h}\bigr) =-\frac{\tilde{h}^{2}}{2}+
\frac{1}{\left(1+\varepsilon^{2}\right)}\frac{\tilde{h}^{4}}{2},  \label{IV:24}
\end{equation}
and
\begin{equation}
\tilde{L}_{3}=-4\pi v^{2}\int\limits_{0}^{\infty }\left[ \frac{1}{8}\frac{
\tilde{h}^{6}\left( \varrho \right)}{m^{4}\left(1+\varepsilon^{2}\right)}
\right]\varrho^{2}d\varrho.                                       \label{IV:25}
\end{equation}
Note that both $\tilde{L}_{1}$  and  $\tilde{L}_{3}$ do not depend on $\omega$.

In the thick-wall regime, the parameter $m_{\omega}=(m^{2} - \omega^{2})^{1/2}$
is small, and therefore we can  neglect  the  second term in Eq.~(\ref{IV:22}).
Then  by  analogy  with  Eqs.~(\ref{IV:18})  and  (\ref{IV:19}),  we obtain the
expressions  of  the  Noether  charge  and  the  energy  of  the soliton in the
thick-wall regime
\begin{equation}
Q\left( \omega \right) =\frac{dL}{d\omega }=\omega \left( m^{2}-\omega
^{2}\right) ^{-1/2}\bigl\vert \tilde{L}_{1}\bigr\vert             \label{IV:26}
\end{equation}
and
\begin{equation}
E\left(\omega \right)=\omega\frac{dL}{d\omega}-L\left(\omega\right)
=m^{2}\left(m^{2}-\omega^{2}\right)^{-1/2}
\bigl\vert \tilde{L}_{1} \bigr\vert.                              \label{IV:27}
\end{equation}
Using Eqs.~(\ref{IV:26}) and (\ref{IV:27}),  we  find  that  in  the thick-wall
approximation, the derivative $dE/dQ = \omega$, and the ratio
\begin{equation}
\frac{E}{Q}=\frac{m^{2}}{\omega }.                                \label{IV:28}
\end{equation}

Thus, in the thick-wall regime $\left\vert\omega \right\vert\rightarrow m$, the
amplitude of the  ansatz   function   $h(r)$   tends  to  zero  proportional to
$m_{\omega} \approx (2m)^{1/2}(m - \left\vert\omega\right\vert)^{1/2}$.
In addition, due to the combination $m_{\omega} r$  in the argument, the ansatz
function $h(r) = m_{\omega}m^{-1}\tilde{h}(m_{\omega} r)$ spreads out in space.
In   contrast,  $u(r)$   and  $1 - u(r)$   remain   of   the  order  of  unity.
Both the energy and the magnitude of the Noether  charge  increase indefinitely
$\propto (m - \left\vert\omega\right\vert)^{-1/2}$  as  $\left\vert\omega\right
\vert \rightarrow m$.
This growth, however, is  significantly  weaker than in the thin-wall regime in
which  it  is   $\propto  \left(\left\vert \omega \right\vert  -  \omega_{\min}
\right)^{-3}$.

\section{Numerical results}                                       \label{sec:V}

The system  of  differential  equations  (\ref{III:9b}) and (\ref{III:9c}) with
the boundary conditions
\begin{eqnarray}
u(0) &=&1,\quad \underset{r\rightarrow \infty }{\lim }u\left( r\right) = 1,
                                                                    \nonumber
 \\
h(0) &=&0,\quad \underset{r\rightarrow \infty }{\lim }h\left( r\right) = 0,
                                                                    \label{V:1}
\end{eqnarray}
is a Dirichlet  boundary  value   problem   on  the  semi-infinite  interval $r
\in \left[0, \infty \right)$,  which  can  be solved only by numerical methods.
To solve this bvp problem, we used the  numerical  methods  of the {\sc{Maple}}
package \cite{maple_2022}.
To check the correctness  of  the  numerical  solutions  obtained,  we used the
virial relation (\ref{III:36}) and  the  differential  relation (\ref{III:39}).

The bvp problem  depends  on  the  five parameters $g$, $\omega$, $m$, $v$, and
$\varepsilon$, while the  ansatz functions $u(r)$ and $h(r)$ are dimensionless.
Eqs.~(\ref{III:9b})  and  (\ref{III:9c}) tell  us   that   these  two functions
depend on four dimensionless variables  and  can be written as $u(\rho, \tilde{
\omega}^{2},\kappa^{2},\varepsilon)$ and $h(\rho,\tilde{\omega}^{2},\kappa^{2},
\varepsilon)$,   where   $\rho = mr$, $\tilde{\omega}=\omega /m$, and $\kappa =
m_{V}/m = \sqrt{2} g v/m$.
Using this information and Eqs.~(\ref{III:13}) and (\ref{III:15}), we find that
the energy and the Noether charge of the soliton can be written as
\begin{equation}
E=\frac{v}{g}\tilde{E}\left( \tilde{\omega }^{2},\kappa
^{2},\varepsilon \right)                                            \label{V:2}
\end{equation}
and
\begin{equation}
Q=\frac{v}{mg}\tilde{Q}\left( \tilde{\omega },
\kappa ^{2},\varepsilon \right)                                     \label{V:3}
\end{equation}
respectively.
Note that the energy (Noether charge)  is  an  even (odd) function of the phase
frequency.

\begin{figure}[tbp]
\begin{center}
\includegraphics[width=0.55\textwidth]{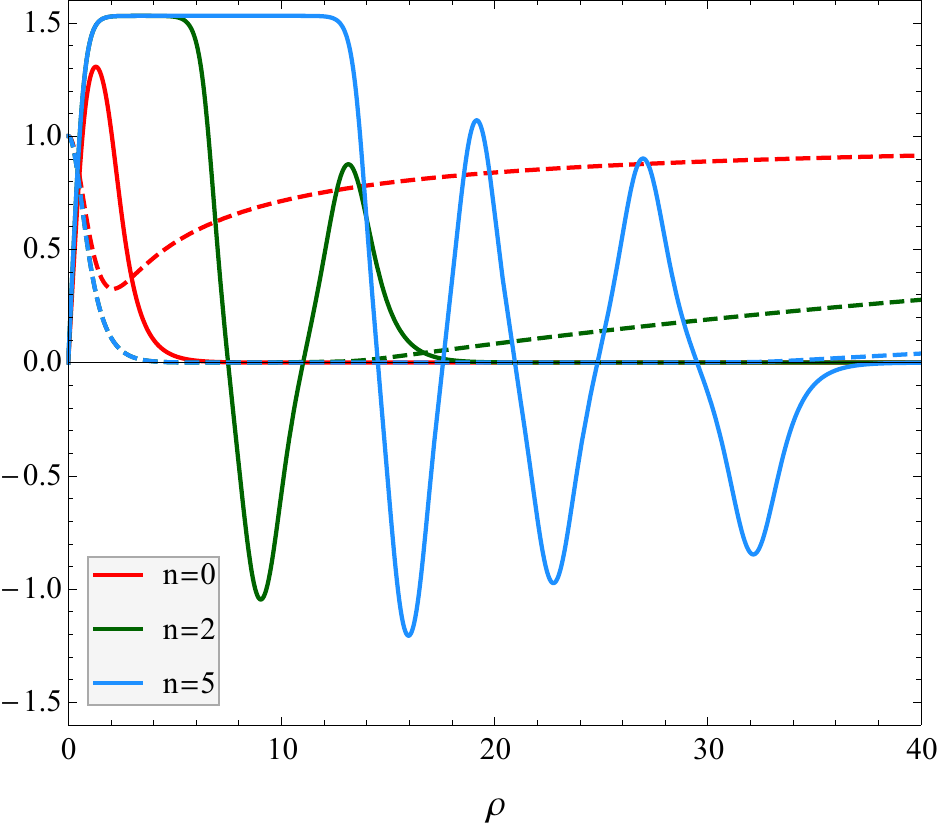}
\caption{\label{fig1}     The  ansatz  functions  $h(\rho)$  (solid curves) and
$u(\rho)$ (dashed curves) for  the  nodeless ($n = 0$)  and two nodal ($n = 2\;
\text{and}\; 5$) soliton solutions.}
\end{center}
\end{figure}

From the above, it follows that, in essence,  the  bvp  depends nontrivially on
the   three    dimensionless    parameters   $\tilde{\omega}$,   $\kappa$,  and
$\varepsilon$.
Furthermore, we see that the energy  $E$  is   proportional  to the combination
$v/g$, which, in turn,  is  $\propto g^{-2}$,  provided  that  $\kappa$ remains
constant.
Hence, the Lagrangian $L$ and the action $S_{T} = 2 \pi \omega^{-1} L$ are also
$\propto g^{-2}$.
It follows that the semiclassical  regime  corresponds  to the condition $g \ll
1$, which we shall assume to be satisfied.

The behavior of the ansatz  function  $h$  is determined by Eq.~(\ref{III:9c}).
Except for  the  term  $-2 r^{-2} u^{2} h$,  which  decreases  rapidly  with an
increase in $r$, this  equation  coincides  with  the  one  that  describes the
behavior of   the   amplitude   of   the   complex  scalar  field  of  a Q-ball
\cite{coleman_1985, paccetti_2001}.
Furthermore, the sixth-order effective potential $\tilde{U}_{\omega}$  entering
Eq.~(\ref{III:9c}) admits  the  existence  of  a series  of  Q-ball  solutions.
Besides the basic nodeless Q-ball  solution,  the effective potential admits an
infinite sequence of excited  Q-ball  solutions  with  monotonically increasing
number of radial zeros.
Considering the above, we  can assume that the  model  (\ref{II:1}) has a basic
soliton solution and a series of radially excited solutions.

Figure~\ref{fig1} shows the  ansatz  functions  $h(\rho)$ and $u(\rho)$ for the
nodeless and two nodal (excited) soliton solutions.
The curves  in  Fig.~\ref{fig1}  correspond   to  the  dimensionless parameters
$\tilde{\omega} = 0.7$, $\kappa = 1$, and $\varepsilon = 1/3$.
To avoid clutter,   we   limited  ourselves   to   the   two   nodal  solutions
corresponding to $n = 2 \; \text{and}\; 5$.
It follows from Fig.~\ref{fig1}  that  the spatial (radial) size of the soliton
increases rapidly with an increase in the number of nodes $n$.
This applies   to   both   the   central   and   nodal  parts  of  the soliton.
We also note that as the number  of  nodes  increases, the  central part of the
soliton has a behavior specific to the thin-wall regime.
In particular, the  ansatz  function $h(\rho)$ is approximately constant in the
central region (except  for  the  monopole-like  core), and the ansatz function
$u(\rho)$  is close to zero in most of the central and nodal regions.

According to Eq.~(\ref{III:15b}), the ansatz function $u(\rho)$  determines the
chromomagnetic field of the soliton.
In particular, Eq.~(\ref{III:15b}) tells us that the  chromomagnetic  field can
be  divided   into   longitudinal   ($\propto  n^{a} n^{i}$)   and   transverse
($\delta^{a i}-n^{a} n^{i}$) components.
The values of the longitudinal and  transverse components are determined by the
scalar functions $B_{\parallel}\left(r\right)=(u\left(r\right)^{2}-1)/(gr^{2})$
and  $B_{\perp}\left(r\right) = u^{\prime} \left(r\right)/(g r)$, respectively.
In the case of the magnetic monopole, the longitudinal component determines the
long-range ($\propto r^{-2}$) part  of  the magnetic field corresponding to  an
unbroken Abelian subgroup of a non-Abelian gauge group.
At   the   same   time,   the   transverse   component   is   the   short-range
($\propto \exp(-m_{V} r)$) part  of the monopole's magnetic field corresponding
to the broken part of the non-Abelian gauge group.

\begin{figure}[tbp]
\begin{center}
\includegraphics[width=0.55\textwidth]{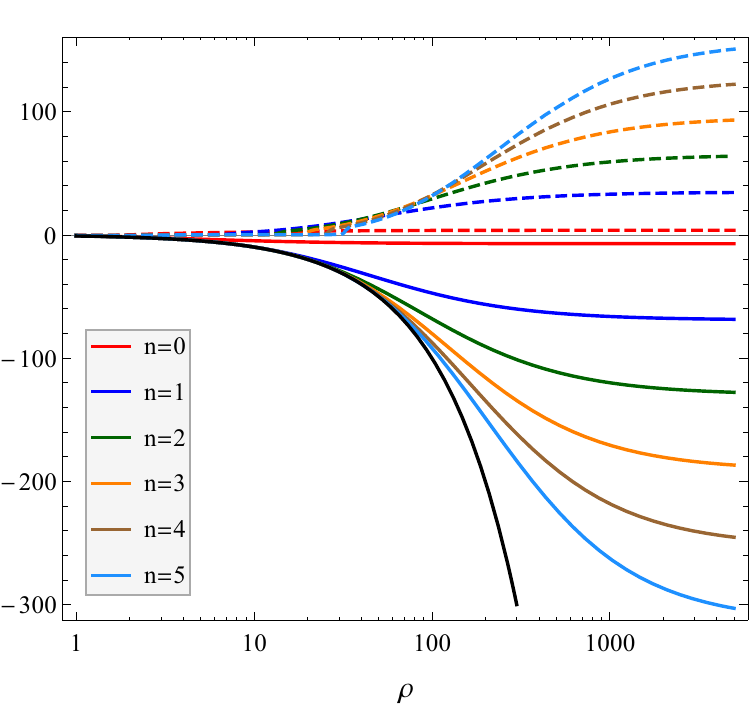}
\caption{\label{fig2}  The curves $\rho^{3}\tilde{B}_{\parallel}(\rho)$ (solid)
and $\rho^{3}\tilde{B}_{\perp}(\rho)$ (dashed)  for  the nodeless and the first
five  nodal  soliton  solutions.       The  black  solid  curve  corresponds to
$\rho^{3}  \tilde{B}_{\text{m}}(\rho)$,   where   $\tilde{B}_{\text{m}}(\rho) =
-\rho^{-2}$    is    the    long-range    Abelian    magnetic    field   of the
't Hooft-Polyakov monopole.}
\end{center}
\end{figure}

The isovector scalar field of the monopole is different from zero at large $r$,
and therefore the gauge symmetry  is  broken  there, and the transverse part of
the monopole's  magnetic  field  is  exponentially  suppressed  compared to the
longitudinal part.
In  contrast,  the  isovector  scalar  field   of   the  nontopological soliton
considered here vanishes exponentially at large $r$.
As a result,  the  gauge  symmetry  is  unbroken  at  large  $r$,  and both the
longitudinal and transverse parts of the chromomagnetic  field   are long-range
($\propto r^{-3}$).

We  define  the  dimensionless   rescaled   versions   of   $B_{\parallel}$ and
$B_{\perp}$ as $\tilde{B}_{\parallel}=g m^{-2} B_{\parallel} = (u(\rho)^{2}-1)/
\rho^{2}$  and  $\tilde{B}_{\perp}  =  g  m^{-2}  B_{\perp}  =  u'(\rho)/\rho$,
respectively.
Figure~\ref{fig2} shows the   curves  $\rho^{3}\tilde{B}_{\parallel}(\rho)$ and
$\rho^{3}\tilde{B}_{\perp}(\rho)$ for  the  nodeless  and  the first five nodal
soliton solutions on a semilogarithmic scale.
It also shows the curve  $\rho^{3} \tilde{B}_{\text{m}}(\rho)$, where $\tilde{B
}_{\text{m}}(\rho) = -\rho^{-2}$  is  the  long-range Abelian magnetic field of
the 't Hooft-Polyakov monopole.
We see that $\tilde{B}_{\parallel}(\rho)$  and $\tilde{B}_{\perp}(\rho)$ behave
quite differently  in the inner region of the soliton, where the gauge symmetry
is partially broken.
$\tilde{B}_{\parallel}(\rho)$   corresponds   to   the   unbroken  $U(1)$ gauge
subgroup, and  therefore  it  is long-range ($\propto \rho^{-2}$) just like the
Abelian part of the monopole's magnetic field.
In contrast,  similarly  to  the  non-Abelian  part  of the monopole's magnetic
field,  $\tilde{B}_{\perp}(\rho)$  is  exponentially  suppressed  in  the inner
region.

\begin{figure}[tbp]
\begin{center}
\includegraphics[width=0.55\textwidth]{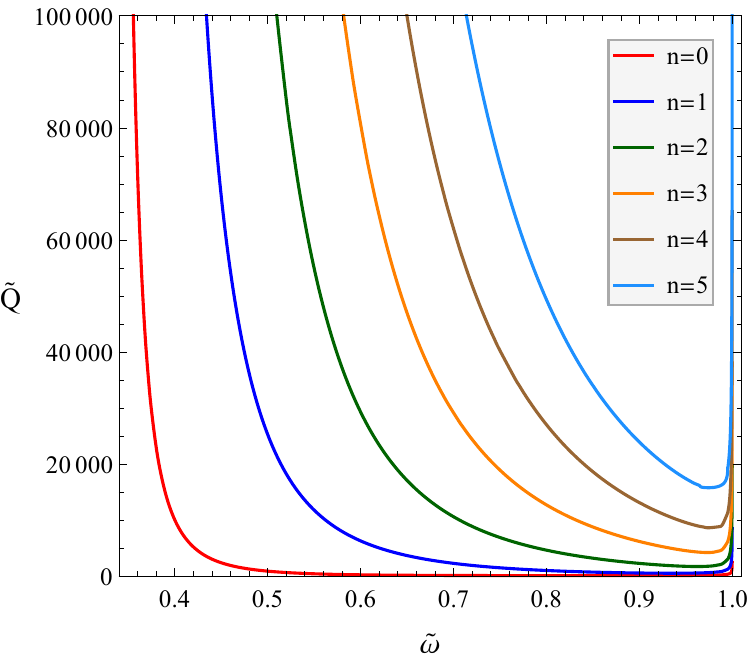}
\caption{\label{fig3}  The curves $\tilde{Q}(\tilde{\omega})$  for the nodeless
and the first five nodal soliton solutions.}
\end{center}
\end{figure}

The situation changes in the external  region, where the isovector scalar field
vanishes and the gauge symmetry is restored.
For this  reason,  both  $\tilde{B}_{\parallel}(\rho)$  and  $\tilde{B}_{\perp}
(\rho)$ are long-range in this region.
As $\rho$ increases, the  long-range character of $\tilde{B}_{\parallel}(\rho)$
changes     from     pole-like     ($\propto   \rho^{-2}$)    to    dipole-like
($\propto \rho^{-3}$).
The behavior   of  $\tilde{B}_{\perp}(\rho)$   also   becomes   dipole-like for
sufficiently large $\rho$.
Note that according  to  Eq.~(\ref{III:28}),  the ratio $\tilde{B}_{\parallel}/
\tilde{B}_{\perp} \rightarrow -2$ as $\rho \rightarrow \infty$.

\begin{figure}[tbp]
\begin{center}
\includegraphics[width=0.55\textwidth]{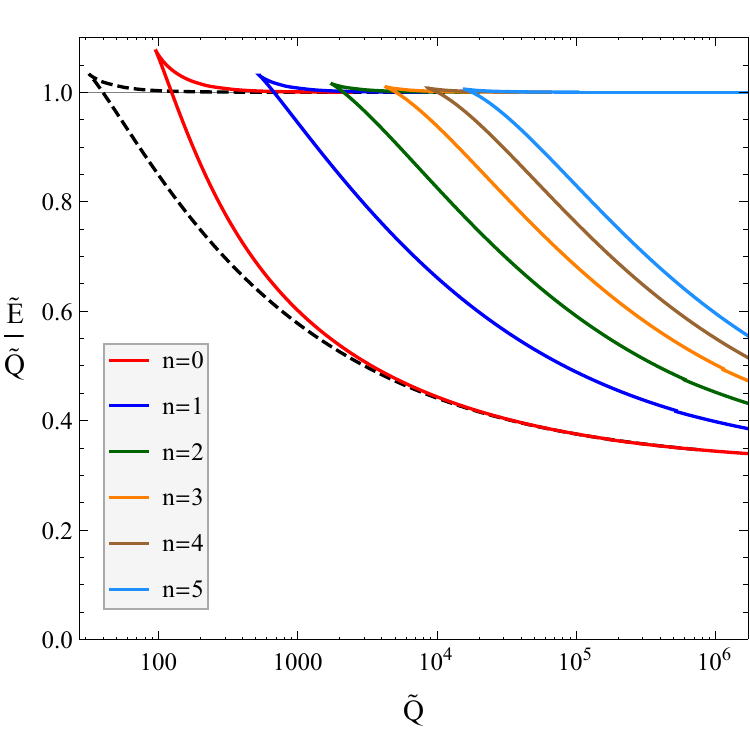}
\caption{\label{fig4}       Dependence  of  the  ratio $\tilde{E}/\tilde{Q}$ on
$\tilde{Q}$.      The solid lines correspond to the nodeless and the first five
nodal soliton solutions.   The black dashed  line  corresponds  to the nodeless
Q-ball solution.}
\end{center}
\end{figure}

\begin{figure}[tbp]
\begin{center}
\includegraphics[width=0.55\textwidth]{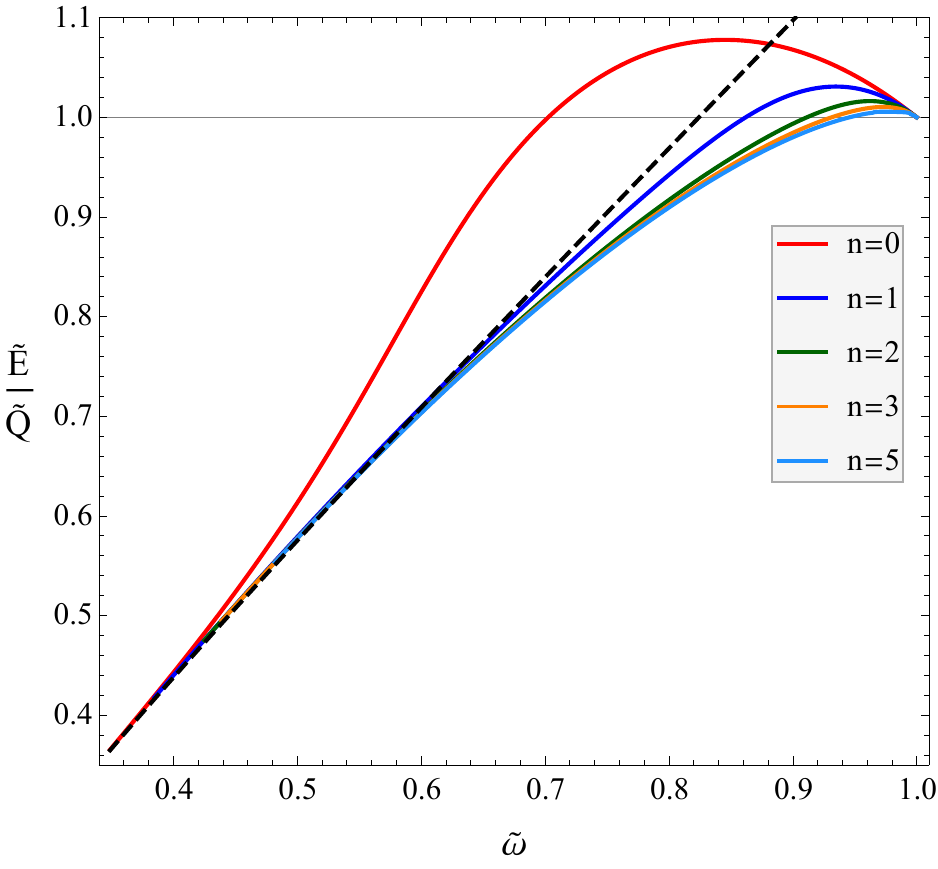}
\caption{\label{fig5}      Dependence  of  the  ratio  $\tilde{E}/\tilde{Q}$ on
$\tilde{\omega}$ for the nodeless  and the first five nodal  soliton solutions.
The curve for $n = 4$ is visually indistinguishable from the curve for $n = 5$.
The  black  dashed  curve  corresponds to Eq.~(\ref{IV:20}).}
\end{center}
\end{figure}

\begin{figure}[tbp]
\begin{center}
\includegraphics[width=0.55\textwidth]{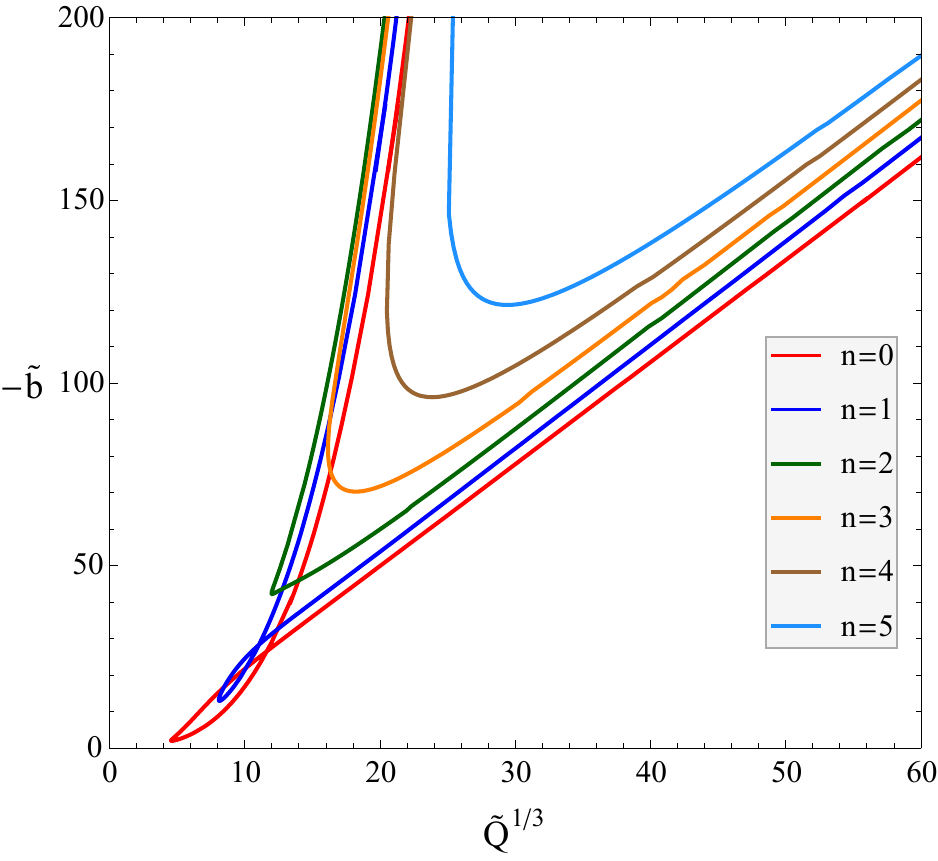}
\caption{\label{fig6} The curves $-\tilde{b}(\tilde{Q}^{1/3})$ for the nodeless
and the first five nodal soliton solutions.}
\end{center}
\end{figure}

Figure~\ref{fig3}  shows   the   dependence  of  the  rescaled  Noether  charge
$\tilde{Q} = m v^{-1} g Q$ on the dimensionless phase frequency $\tilde{\omega}
=m^{-1}\omega$  for the  nodeless  and  the first five nodal soliton solutions.
The  curves  in  Fig.~\ref{fig3} and  subsequent Figs.~\ref{fig4} -- \ref{fig6}
correspond  to  the  dimensionless  parameters  $\kappa = 1$ and $\varepsilon =
1/3$.
We see that $\tilde{Q}$  increases without bound as $\tilde{\omega}$ decreases,
which corresponds to the transition to the thin-wall regime.
It was found numerically that in this  case, $\tilde{Q}(\tilde{\omega}) \propto
(\tilde{\omega} - \tilde{\omega}_{\min})^{-3}$,  where $\tilde{\omega}_{\min} =
\varepsilon (1 + \varepsilon^{2})^{-1/2} = 10^{-1/2}$.
Such a  behavior  of  the  Noether   charge   $\tilde{Q}$   is  consistent with
Eq.~(\ref{IV:18}).
The Noether charges of  the  soliton  solutions also increase indefinitely when
$\tilde{\omega} \rightarrow 1$,  which  corresponds  to  the  transition to the
thick-wall regime.
In this case, the Noether charges $\tilde{Q}(\tilde{\omega}) \propto (1-\tilde{
\omega})^{-1/2}$  , which is consistent with Eq.~(\ref{IV:26}).
It follows from Fig.~\ref{fig3}  that  at a fixed $\tilde{\omega}$, the Noether
charge $\tilde{Q}$  increases  monotonically  with an increase in the number of
nodes $n$, and the rate of increase also grows with $n$.
The behavior of  the  curves  $\tilde{E}(\tilde{\omega})$  is  similar  to that
presented in Fig.~\ref{fig3}.
In particular, the  soliton's  energy  $\tilde{E}  \rightarrow  \infty$ both as
$\tilde{\omega}   \rightarrow   \tilde{\omega}_{\min}$    and   $\tilde{\omega}
\rightarrow 1$.

In addition  to   the   soliton   solutions   under   consideration,  the model
(\ref{II:1}) also  has  a  simple  soliton   solution  described  by the ansatz
$A_{\mu }^{a}=0,\; \phi ^{a} = 2^{-1/2}vh\left( r \right) e^{i \omega t} \left(
0,0,1\right)$ and the boundary conditions $h^{\prime}\left(0 \right)  =  0$ and
$h\left(\infty\right)=0$.
Since the gauge field is completely decoupled  from the scalar isotriplet, this
soliton solution is in essence an ordinary Q-ball.
Figure~\ref{fig4} presents the dependence of the ratio $\tilde{E}/\tilde{Q}$ on
$\tilde{Q}$ for  the  nodeless  and  the  first  five  nodal  soliton solutions
on a semilogarithmic scale.
It also  presents   the   dependence   of  the  ratio  $\tilde{E}/\tilde{Q}$ on
$\tilde{Q}$ for the nodeless Q-ball solution.
A characteristic feature  of  a  curve  in  Fig.~\ref{fig4} is a cuspidal point
that divides the curve into upper and lower branches.
The presence  of  the  cuspidal  point  is  a  typical  sign  of nontopological
solitons \cite{fried_1976, lee_pang_1992}.
It follows  from  Fig.~\ref{fig4}  that  the  Noether  charge  $\tilde{Q}$ of a
soliton reaches the minimum  at this point.
This is because the cuspidal point of a curve in Fig.~\ref{fig4} corresponds to
the  minimum point  of  the  corresponding curve $\tilde{Q}(\tilde{\omega})$ in
Fig.~\ref{fig3}.
Furthermore, Eq.~(\ref{III:39}) tells us  that  the  energy of the soliton also
reaches the minimum at the cuspidal point.

Let   us    consider    the     stability    of    the    soliton    solutions.
The instability of a soliton can be either  classical  (unstable  mode  (modes)
in the spectrum  of  fluctuation)  or  quantum-mechanical  (tunneling process).
It was shown in \cite{lee_pang_1992} that a cuspidal  point  on  a $E(Q)$ curve
means the appearance of an unstable fluctuation  mode  on  the  upper branch of
the curve.
Hence, in Fig.~\ref{fig4}, the upper branches of all  the  curves correspond to
classically unstable solitons.
In addition, it was shown in  \cite{fried_1976} that radial nodes  lead  to the
classical instability of soliton solutions.
It follows that in Fig.~\ref{fig4}, the lower branches of  the curves for which
$n > 0$ also correspond to classically unstable solitons.
Finally, it  was established in \cite{fried_1976} and \cite{paccetti_2001} that
the  stability  criterion for nodeless  nontopological   soliton   solutions is
given by  the condition 
\begin{equation}
\frac{\omega}{Q}\frac{dQ}{d\omega} < 0,                             \label{V:4}
\end{equation}
where $Q\,$ ($\omega$) denotes  the  soliton  Noether charge (phase frequency).
It follows from Fig.~\ref{fig3} that the lower branch of the red solid curve in
Fig.~\ref{fig4} satisfies this condition.
Hence, only   this   branch    corresponds   to   classically  stable solitons.

In Fig.~\ref{fig4}, the red solid curve corresponds to  the  nodeless solitons,
whereas  the   black  dashed   curve   corresponds   to  the  nodeless Q-balls.
We see that for a given $Q$, the  energy  of  the  nodeless Q-ball is less than
the  energy of the nodeless  soliton,  although the difference becomes visually
indistinguishable at large $Q$.
Hence, the classically stable nodeless soliton is quantum-mechanically unstable
to the transition into the nodeless Q-ball.
This instability, however, is not a serious problem.
Indeed,  according to \cite{E_Weinberg}, the  nucleation  rate  per unit volume
is, to an order of magnitude,
\begin{equation}
\Gamma \sim \frac{m^{4}}{g^{4}}e^{-B},                              \label{V:5}
\end{equation}
where $B$ is  the tunneling  exponent, $g$  is  the  gauge  coupling  constant,
$m$  is  the mass of the isovector scalar field $\boldsymbol{\phi}$, and  it is
assumed that the ratio $\kappa = \sqrt{2}gv/m$ is of order unity.

The tunneling exponent $B$ is  the  Euclidean  action  of a field configuration
that interpolates between the initial and final states.
It was shown  in  \cite{E_Weinberg} that, in  the  general  case, the tunneling
exponent is
\begin{equation}
B = \frac{2\pi^2}{g^2} C,                                           \label{V:6}
\end{equation}
where $C$ is of order unity.
Furthermore, classical  solutions  are  meaningful  only  in  the semiclassical
regime, where quantum corrections are small.
In this regime, $g \ll 1$ and $\kappa=\sqrt{2}gv/m$ is  fixed; hence, the gauge
coupling constant $g$ is small.
Therefore, the tunneling  exponent  (\ref{V:6})  is  large,  and the nucleation
rate (\ref{V:5}) is exponentially suppressed.
As a result,  the  mean lifetime of the nodeless soliton can reach cosmological
scales.

Let us  discuss  some  consequences  of  the  classical  instability of soliton
solutions.
One such consequence is the decay of  an  unstable soliton into a multiparticle
state.
In Fig.~\ref{fig4}, each curve intersects the line $\tilde{E}/\tilde{Q} = 1$ at
a single point on the lower branch.
In addition, the upper branch  of each curve approaches unity from above in the
limit $\tilde{Q} \rightarrow \infty$.
The ratio  $\tilde{E}/\tilde{Q}$  serves  as  a  criterion  for  stability of a
soliton with Noether  charge  $\tilde{Q}$  against  decay  into a multiparticle
state with the same charge.
In the  dimensionless  units   employed   here,   the   rest  energy  of such a
multiparticle state is $\tilde{Q}$, implying that the minimum value of $\tilde{
E}/\tilde{Q}$ is equal to unity.
Energy considerations then imply that the  soliton is stable (unstable) against
decay into the multiparticle state when the ratio $\tilde{E}/\tilde{Q}$ is less
(greater) than unity.
Accordingly, classically unstable solutions corresponding to the upper branches
of the $\tilde{E}/\tilde{Q}$  curves  undergo  decay into multiparticle states.
The same conclusion  applies  to  classically  unstable  solutions of the lower
branches for which $\tilde{E}/\tilde{Q} > 1$.

We now turn to classically unstable  solutions  for  which  the ratio $\tilde{E
}/\tilde{Q} < 1$.
They correspond to the lower branches  of  the solid curves in Fig.~\ref{fig4},
except for the red curve.
For  these  solutions,  decay  into  a  multiparticle  state  is  energetically
forbidden.
Instead, they transition into either a stable nodeless Q-ball (the lower branch
of the black dashed curve) or a  metastable nodeless soliton  (the lower branch
of the red solid curve).
The energy and Noether charge  released  during  this transition are emitted in
the form of particle radiation.

Figure~\ref{fig5} shows the dependence  of  the  ratio $\tilde{E}/\tilde{Q}$ on
the phase frequency $\tilde{\omega}$ for the nodeless  and the first five nodal
soliton solutions.
It also shows  a  dashed curve that corresponds  to  the  thin-wall  regime and
is described by Eq.~(\ref{IV:20}).
The cuspidal points of the curves in Fig.~\ref{fig4} correspond  to the maximum
points  of the curves in Fig.~\ref{fig5}.
As  in  Fig.~\ref{fig4}, each  curve  in  Fig.~\ref{fig5}  intersects  the line
$\tilde{E}/\tilde{Q} = 1$ at a single point, and all these curves tend to unity
from above as $\tilde{\omega} \rightarrow 1$.
This corresponds   to   the   thick-wall   regime   and   is  in agreement with
Eq.~(\ref{IV:28}).
As the  phase  frequency  $\tilde{\omega}$  decreases,  the  soliton enters the
thin-wall regime in  which  the  ratio  $\tilde{E}/\tilde{Q}$  is determined by
Eq.~(\ref{IV:20}).
It follows from Fig.~\ref{fig5}  that in this regime, all the curves merge with
the dashed curve.
Hence, we conclude  that  as  $\tilde{\omega}$  decreases,  the behavior of the
solitons is well described within the thin-wall approximation.

A characteristic feature  of  the  studied  solitons  is  the long-range dipole
($\propto r^{-3}$) chromomagnetic field.
Eq.~(\ref{III:27})  tells   us   that   the   soliton's  chromomagnetic  dipole
moment  is  determined  by  the  parameter  $b$,   which   also  determines the
long-range  ($\propto r^{-1}$)  asymptotics  of  the  ansatz  function  $u(r)$.
We define the dimensionless version of $b$ by $\tilde{b} = m b$ and recall that
$b$ is negative.
In Fig.~\ref{fig6}, we  can see the graphs of $-\tilde{b}(\tilde{Q}^{1/3})$ for
the nodeless and the first five nodal soliton solutions.
The choice  of  $\tilde{Q}^{1/3}$   as  an  argument  is  because  according to
Eqs.~(\ref{IV:16}) and (\ref{IV:18}), the  effective  radius $R$ of the soliton
is $\propto Q^{1/3}$ in the thin-wall regime.

Each curve in Fig.~\ref{fig6} has two branches  which  have a self-intersection
point  when  the  number  of  nodes $n = 0\; \text{or}\; 1$.
As $\tilde{Q}$  increases,  the  lower  (upper)  branch  goes  to the thin-wall
(thick-wall)   regime   in    which   $-\tilde{b}$    increases   indefinitely.
It was found numerically  that  the  lower  branches  of the curves $-\tilde{b}
(\tilde{Q}^{1/3})$    are   asymptotically   linear    in    $\tilde{Q}^{1/3}$.
It follows that in the thin-wall  regime, $b$  is proportional to the effective
soliton radius $R$.

In contrast,   the   upper   branches   of   $-\tilde{b}(\tilde{Q}^{1/3})$  are
asymptotically  cubic  in   $\tilde{Q}^{1/3}$,  i.e.  they  are  asymptotically
$\propto \tilde{Q}$ in the thick-wall regime.
In this regime,  the   soliton  spreads  unboundedly   through  space,  and its
boundaries become blurred.
In  this  case,   according   to   Subsection~\ref{subsec:IVB},  the  parameter
$m_{\omega }^{-1}=\left( m^{2} - \omega ^{2}\right) ^{-1/2}$ can be taken as an
estimate of the size of the soliton.
At the same time, Eq.~(\ref{IV:26}) tells  us  that  in  the thick-wall regime,
the Noether charge $Q$ is also proportional to $m_{\omega}^{-1}$.
Summarising  these  facts,  we  conclude  that  in  the  thick-wall regime, $b$
is proportional to the  effective soliton radius.
We see that in  both  extreme  regimes,  the  chromomagnetic  dipole  moment is
proportional to the  soliton's  linear  size,  which  was to be expected on the
basis of dimensional considerations.

\section{Conclusion}                                             \label{sec:VI}

We have investigated  a  nontopological  soliton  of a non-Abelian gauge model.
This model has similarities  with  the  well-known $SU(2)$ Geogri-Glashow model
while differing from it in three important aspects.
First, the real scalar isotriplet of  the Georgi-Glashow model is replaced by a
complex scalar isotriplet.
The transition to the complex scalar isotriplet  results in a new global $U(1)$
symmetry and a corresponding Noether charge.
Second, unlike the Georgi-Glashow model, the  potential of the considered model
has a global minimum at zero value of the scalar isotriplet, and  therefore the
$SU(2)$  gauge   symmetry  is  not  spontaneously  broken.
Third, the self-interaction  potential  of  the  Georgi-Glashow model is of the
fourth  order  in  the  field,  and  therefore  the  model  is  renormalizable.
At the same time, the self-interaction  potential of the considered model is of
the sixth order in the field,  and  its  form  coincides  with that used in the
study of Q-balls.

The above differences make possible the existence of  a  nontopological soliton
in the framework of the model (\ref{II:1}).
Many  properties   of   this   soliton   coincide   with  those  of  a  Q-ball.
These include a cuspidal point  on  the $E(Q)$  curve, the  presence of the two
(thin-wall and thick-wall) limiting  regimes,  and  the  existence  of radially
excited nodal solutions.
A characteristic property  of  the  soliton  bringing  it  closer to a magnetic
monopole is its spherically symmetric magnetic field.
In the thin-wall regime, the magnetic field of the  soliton in the inner region
practically coincides with that of a magnetic monopole.
In particular, in the inner  region  of the soliton, the  Abelian  component of
the  magnetic   field   $B_{i} \approx n_{i}/\left(g r\right)^{2}$,  while  the
non-Abelian component is suppressed exponentially.

The situation changes in the external  region,  where the scalar field vanishes
and the non-Abelian gauge symmetry is restored.
In this region, the Abelian component of the magnetic field has the asymptotics
of dipole ($\propto r^{-3}$) type,  which  means  zero  magnetic  charge of the
soliton.
Moreover, in this region, the  other components of the magnetic field also have
long-range dipole ($\propto r^{-3}$) asymptotics.
Thus, at large  distances from the  center,  the soliton possesses a long-range
chromomagnetic field of dipole type.
For the two (thin-wall  and  thick-wall)  limiting  regimes, the chromomagnetic
dipole moment of the soliton is proportional to its linear size.

We can say that the nontopological soliton  is composed of a monopole-like core
surrounded by a Q-ball-like shell.
This shell completely shields the magnetic charge of the core, which results in
the dipole ($\propto r^{-3}$) magnetic field of the soliton at a large distance
from the center.
Just like the usual Q-ball, the size  of  the  shell  increases indefinitely as
$\tilde{\omega} \rightarrow \tilde{\omega}_{\min}$ (thin-wall regime).

Having  a long-range chromomagnetic field, the soliton nevertheless possesses a
zero electric field.
This is because in  the  model (\ref{II:1}),  any  spherically  symmetric field
configurations  with  a non-zero  electric  field   will have infinite  energy.
In contrast,  there  exist  electrically  charged  magnetic  monopoles  (dyons)
which have finite energy \cite{jz_1975}.
In this connection it is worth  noting  that  at large distances from the dyon,
the non-Abelian  gauge  symmetry  is  broken  except  for  the  electromagnetic
Abelian $U(1)$ subgroup.
Hence,  the  long-range   electric    field    of    the    dyon   is  Abelian.
However, the non-Abelian  gauge  symmetry  is  not  broken  at  large distances
from the soliton of the model (\ref{II:1}).
It follows that if the soliton could have a long-range  electric field it would
be non-Abelian (chromoelectric) like its chromomagnetic field.

It is worth noting that there  are  also  no  magnetic  monopoles (chromodyons)
possessing   a  long-range  chromoelectric  field   \cite{abouelsaood_npb_1983,
 abouelsaood_plb_1983,     balan_1983,     nelson_1983,    nelson_manohar_1983,
  nelson_coleman_1984}.
The reason is that the long-range tails of  the  non-Abelian components  of the
fields produce a topological obstruction that  makes it impossible  to  find  a
basis for an unbroken  non-Abelian  subgroup  that is smooth over all of space.
As a result, it is impossible to define  normalizable  zero  modes  which would
correspond to some generators of the unbroken gauge non-Abelian subgroup.

The nontopological   soliton   considered   here   is   related  to  the gauged
monopole-bubble solution of \cite{kim_plb_1999}.
The gauged monopole-bubble contributes to the decay amplitude of a false vacuum
in the high-temperature limit.
Unlike the model (\ref{II:1}),  the  model  considered  in  \cite{kim_plb_1999}
contains a real isovector scalar field, so that  the  conserved  Noether charge
(\ref{III:15}) is absent.
In addition,  the   true  vacuum   manifold   of  this  model  is  a two-sphere
$\left\vert \boldsymbol{\phi} \right\vert = \phi_{\text{vac}}$  rather than the
point $\boldsymbol{\phi} = 0$ as it is in the model (\ref{II:1}).
We can  say  that  the  gauged  monopole-bubble  of \cite{kim_plb_1999} and the
nontopological soliton  of  the model (\ref{II:1}) are related similarly to the
three-dimensional  Euclidean   bounce   and   the  $(3+1)$  dimensional Q-ball.
Note also that a nontopological soliton  exists  in  the  non-gauge  version of
the model (\ref{II:1}), which describes  a  self-interacting  isovector complex
scalar field.
This  soliton,   called    Q-crust,    was   described   in  \cite{sakai_2011}.

In this paper, we considered  the  simplest non-Abelian gauge model in  which a
Q-ball-like soliton solution exists.
Similar  to  the 't Hooft-Polyakov   monopole,  this  soliton  solution  can be
embedded in a gauge group larger than $SU(2)$.
To do  this, it  is  sufficient  to  complexify  the  adjoint  Higgs  field and
choose  any of the $SU(2)$ subgroups  of  the  gauge group.
The choice of the $SU(2)$ subgroup  determines  the  embedding according to the
formulas given in \cite{weinberg_2007}.
The soliton     solution    can      be   generalized     in     another   way.
For this, the  global $U(1)$  symmetry  (\ref{II:7})  must  be  gauged using an
Abelian gauge field.
In this case, the soliton will have an Abelian electric field and a non-Abelian
chromomagnetic field.

\section*{Acknowledgements}

This  work  was  supported   by   the   Russian  Science  Foundation,  grant No
23-11-00002-Ext.


\begin{thebibliography}{99}

\bibitem{Manton}
N.~Manton, P.~Sutclffe, Topological Solitons, Cambridge University Press,
Cambridge, 2004.

\bibitem{E_Weinberg}
E.J. Weinberg, Classical Solutions in Quantum Field Theory: Solitons and
Instantons in High Energy Physics, Cambridge University Press, Cambridge, 2012.

\bibitem{lee_pang_1992}
T.D.~Lee, Y.~Pang, Phys. Rep. 221 (1992) 251.

\bibitem{hooft_74}
G.~’t Hooft, Nucl. Phys. B 79 (1974) 276.

\bibitem{polyakov_74}
A.M.~Polyakov, JETP Lett. 20 (1974) 194.

\bibitem{GG_1972}
H.~Georgi, S.L.~Glashow, Phys. Rev. Lett. 28 (1972) 1494.

\bibitem{kumar_2010}
B.~Kumar, M.B.~Paranjape, U.A.~Yajnik, Phys. Rev. D 82 (2010) 025022.

\bibitem{paranjape_2024}
M.B.~Paranjape, Y.~Saxena, Phys. Rev. D 110 (2024) 025005.

\bibitem{steinhardt_prd_1981}
P.J.~Steinhardt, Phys. Rev. D 24 (1981) 842.

\bibitem{steinhardt_npb_1981}
P.J.~Steinhardt, Nucl. Phys. B 190 (1981) 583.

\bibitem{hosotani_prd_1983}
Y.~Hosotani, Phys. Rev. D 27 (1983) 789.

\bibitem{agrawal_2022}
P.~Agrawal, M.~Nee, SciPost Phys. 13 (2022) 049.

\bibitem{kim_plb_1999}
Y.~Kim, S.~Lee, K.~Maeda, N.~Sakai, Phys. Lett. B 452 (1999) 214.

\bibitem{bai_2022}
Y.~Bai, S.~Lu, N.~Orlofsky, JHEP 01 (2022) 109.

\bibitem{coleman_1977}
S.~Coleman, Phys. Rev. D 15 (1977) 2929.

\bibitem{coleman_1985}
S.~Coleman, Nucl. Phys. B 262 (1985) 263.

\bibitem{gunion_1990}
J. F. Gunion, R. Vega, J. Wudka, Phys. Rev. D 42 (1990) 1673.

\bibitem{rizzo_1991}
T.G. Rizzo, Mod. Phys. Lett. A 6 (1991) 1961.

\bibitem{espinosa_1992}
J.R. Espinosa, M. Quiros, Nucl. Phys. B 384 (1992) 113.

\bibitem{dey_2009}
P. Dey, A. Kundu, B. Mukhopadhyaya, J. Phys. G 36 (2009) 025002.

\bibitem{affleck_1985}
I. Affleck, M. Dine, Nucl. Phys. B 249 (1985) 361.

\bibitem{enqvist_1998}
K. Enqvist, J. McDonald, Phys. Lett. B 425 (1998) 309.

\bibitem{enqvist_1999}
K. Enqvist, J. McDonald, Nucl. Phys. B 538 (1999) 321.

\bibitem{kasuya_2000}
S. Kasuya, M. Kawasaki, Phys. Rev. D 62 (2000) 023512.

\bibitem{kusenko_1998}
A. Kusenko, M. Shaposhnikov, Phys. Lett. B 418 (1998) 46.

\bibitem{kusenko_2009}
I.M. Shoemaker, A. Kusenko, Phys. Rev. D 80 (2009) 075021.

\bibitem{fried_1976}
R.~Friedberg, T.D. Lee, A.~Sirlin, Phys. Rev. D 13 (1976) 2739.

\bibitem{loginov_plb_2021}
A.{\relax Yu}. Loginov, Phys. Lett. B 822 (2021) 136662.

\bibitem{loginov_plb_2024}
A.{\relax Yu}. Loginov, Phys. Lett. B 855 (2024) 138840.

\bibitem{maple_2022}
Maple 2022.1, Maplesoft, a division of Waterloo Maple Inc., Waterloo, Ontario.

\bibitem{paccetti_2001}
F. Paccetti-Correia, M.G.~Schmidt, Eur. Phys. J. C 21 (2001) 181.

\bibitem{jz_1975}
B.~Julia, A.~Zee, Phys. Rev. D 11 (1975) 2227.

\bibitem{abouelsaood_npb_1983}
A.~Abouelsaood, Nucl. Phys. B 226 (1983) 309.

\bibitem{abouelsaood_plb_1983}
A.~Abouelsaood, Phys. Lett. B 125 (1983) 467.

\bibitem{balan_1983}
A.~Balachandran, G.~Marmo, N.~Mukunda, J.~Nilsson,
E.~Sudarshan, F.~Zaccaria, Phys. Rev. Lett. 50 (1983) 1553.

\bibitem{nelson_1983}
P.~Nelson, Phys. Rev. Lett. 50 (1983) 939.

\bibitem{nelson_manohar_1983}
P.~Nelson, A.~Manohar, Phys. Rev. Lett. 50 (1983) 943.

\bibitem{nelson_coleman_1984}
P.~Nelson, S.~Coleman, Nucl. Phys. B 237 (1984) 1.

\bibitem{sakai_2011}
N. Sakai, H. Ishihara, K. Nakao, Phys. Rev. D 84 (2011) 105022.

\bibitem{weinberg_2007}
E.J. Weinberg, P. Yi, Phys. Rep. 438 (2007) 65.

\end{thebibliography}
\end{document}